	\documentclass{ifacconf}

	\usepackage{subfigure}
	
	\newcommand{\R}{\mathbb{R}}

	\usepackage[tracking=true]{microtype}
	\usepackage{tikz}
	\usepackage{float}  
	
	\usepackage{algorithmicx}
	\usepackage{algorithm}
	\usepackage{algpseudocode}

	\usepackage{epsfig}
	\usepackage{multirow}

	\usepackage{amssymb}
	\usepackage{amsmath}
	\usepackage{bm}
	\usepackage{cite}
	\usepackage{breqn}
	\usepackage{float}
	\usepackage{mathrsfs}
	\usepackage{mathtools}
	\usepackage{amsfonts}
	\usepackage{graphicx}
	\usepackage{algorithm, algpseudocode}
	\usepackage{subfigure}

	\usepackage{xcolor}
	\usepackage{amsfonts}
	\usepackage{relsize}
	\usepackage{lmodern}
	\usepackage{slantsc}
	\usepackage[mathscr]{eucal}
	\usepackage{enumerate}
	\usepackage[final]{pdfpages}
	\usepackage{stfloats}
	\usepackage{blindtext}
	
	\usepackage{caption}
	\usepackage{balance}
	\usepackage{flushend}
	\usepackage{epsfig}

	\usepackage{multirow}

	\usepackage{amssymb}
	\usepackage{amsmath}
	\usepackage{bm}
	
	\usepackage{cite}
	
	\usepackage{breqn}
	\usepackage{float}
	\usepackage{mathrsfs}
	\usepackage{mathtools}
	\usepackage{amsfonts}
	\usepackage{graphicx}

	\usepackage{algorithm, algpseudocode}
	
	\usepackage{subfigure}
	\usepackage{xcolor}
	\usepackage{amsfonts}
	\usepackage{relsize}
	\usepackage{lmodern}
	\usepackage{slantsc}
	\usepackage[mathscr]{eucal}
	
	\usepackage{dsfont}
	\usepackage{bbm}
	\usepackage{bm}
	
	\usepackage{caption}
	\usepackage{tabularx, booktabs}
	\usepackage{adjustbox}
	
	\usepackage{xcolor}
	
	\usepackage{colortbl}
	\usepackage{lipsum}
	\usepackage{scalerel}
	\usepackage{stackengine}
	\usepackage{tabularx,ragged2e}
	\usepackage{booktabs}

	\usepackage[utf8]{inputenc}
	\usepackage{array}
	\usepackage{makecell}
	
	\usepackage{enumitem}
	\usepackage{makecell}
	\usepackage{multicol}
	\usepackage{scalerel}
	
	\usepackage[font={small}]{caption}

	\usepackage{graphicx}      
	\usepackage{natbib}        
	\usepackage{amssymb,amsmath,mathrsfs}
	\usepackage{comment}

	\usepackage{amssymb,amsmath,mathrsfs}
	\usepackage{comment}
	
	\usepackage{flushend} 
	\usepackage{balance} 

	\usepackage{algorithm}
	\usepackage{algpseudocode}

	\usepackage[hoptionsi]{overpic}
	\usepackage{bbm}

	\usepackage{algorithmicx}
	\usepackage{algorithm}
	\usepackage{algpseudocode}

	\usepackage{epsfig}
	\usepackage{multirow}

	\usepackage{bm}
	
	\usepackage{float}
	\usepackage{mathrsfs}
	\usepackage{mathtools}
	\usepackage{amsfonts}
	\usepackage{graphicx}
	\usepackage{algorithm, algpseudocode}

	\usepackage{amsfonts}
	\usepackage{relsize}
	\usepackage{lmodern}
	\usepackage{slantsc}
	\usepackage[mathscr]{eucal}
	\usepackage{enumerate}
	\usepackage[final]{pdfpages}
	\usepackage{stfloats}
	\usepackage{blindtext}
	
	\usepackage{caption}
	\usepackage{balance}
	\usepackage{flushend}
	\usepackage{epsfig}

	\usepackage{graphicx}

	\usepackage{algorithm, algpseudocode}

\begin{document}
	\begin{frontmatter}
	
	\title{Distributed Switching Model Predictive Control Meets Koopman Operator for Dynamic Obstacle Avoidance}

	
	
	\author[First]{Ali Azarbahram} 
	\author[Second]{Chrystian Pool Yuca Huanca} 
	\author[Second]{Gian Paolo Incremona} 
	\author[Second]{Patrizio Colaneri}

	\address[First]{The Department of Electrical Engineering, Chalmers University of
	Technology, Gothenburg, 412 96, Sweden. (e-mail: ali.azarbahram@chalmers.se).}
	
	\address[Second]{The Dipartimento di Elettronica, Informazione
	 e Bioingegneria, Politecnico di Milano, 20133 Milan, Italy. (e-mails: \{chrystianpool.yuca, gianpaolo.incremona, and patrizio.colaneri\}@polimi.it).}

	\thanks{This work has been supported by the Italian Ministry of Enterprises and Made in Italy in the framework of the project 4DDS (4D Drone Swarms) under grant no. F/310097/01-04/X56.}
	   
	\begin{abstract}                
	This paper introduces a Koopman-enhanced distributed switched model predictive control (SMPC) framework for safe and scalable navigation of quadrotor unmanned aerial vehicles (UAVs) in dynamic environments with moving obstacles. The proposed method integrates switched motion modes and data-driven prediction to enable real-time, collision-free coordination. A localized Koopman operator approximates nonlinear obstacle dynamics as linear models based on online measurements, enabling accurate trajectory forecasting. These predictions are embedded into a distributed SMPC structure, where each UAV makes autonomous decisions using local and cluster-based information. This computationally efficient architecture is particularly promising for applications in surface transportation, including coordinated vehicle flows, shared infrastructure with pedestrians or cyclists, and urban UAV traffic. Simulation results demonstrate reliable formation control and real-time obstacle avoidance, highlighting the framework’s broad relevance for intelligent and cooperative mobility systems.
	\end{abstract}
	
	\begin{keyword}
	Koopman operator, switched model predictive control, distributed control, dynamic obstacle avoidance, quadrotor UAVs.
	\end{keyword}
	
	\end{frontmatter}
	
	

	\section{Introduction}
	\label{Sec. 1}
	
	Cooperative unmanned aerial vehicle (UAV) navigation is increasingly vital in intelligent mobility systems, where safe interaction with other vehicles, infrastructure, and dynamic road users is paramount \citep{scherer2008flying}. Multi-UAV coordination enhances situational awareness and coverage \citep{michael2011cooperative}, but introduces safety and scalability challenges \citep{ponda2010decentralized, zhou2015virtual}. However,  trajectory planning under actuator and collision constraints must remain computationally feasible for real-time deployment \citep{saccani2022multitrajectory, tang2018autonomous}. Model predictive control (MPC) is widely used for such tasks \citep{mayne2005robust}, while onboard sensing or direct avoidance is often required in partially unknown environments \citep{liu2017planning, hrabar2011reactive}.
	
	MPC’s predictive and constraint-handling capabilities make it attractive for real-time control \citep{garcia1989model, kouvaritakis2016model}, with switched MPC providing further efficiency by restricting the control to a predefined set of motion primitives \citep{bemporad1999control, dsmpc}. These features align well with the challenges of urban UAV navigation operating in proximity to infrastructure or pedestrians.
	
	Obstacle avoidance in dynamic environments demands both rapid response and reliable motion forecasting. Recent MPC-based methods have incorporated probabilistic and learning-based models for obstacle behavior \citep{li2023moving, olcay2024dynamic, wei2022moving}. Despite this progress, few efforts address online modeling of unknown dynamics for embedded control. This motivates the use of the Koopman operator, which linearly embeds nonlinear systems into higher-dimensional spaces, enabling efficient data-driven prediction \citep{williams2015data}. 

    The Koopman operator has emerged as a powerful tool for bridging nonlinear dynamics and linear control frameworks. A comprehensive treatment of its theoretical underpinnings and relevance to systems and control is provided in \citep{mauroy2020koopman}, outlining its use in estimation, observability, and feedback design. The approach is utilized for controlled dynamical systems in \citep{korda2018linear}, where data-driven linear predictors are constructed in lifted spaces, enabling integration into efficient MPC schemes. The broader utility of the Koopman framework in forecasting, system identification, and controller synthesis is also surveyed in \citep{mezic2021koopman}. Koopman-based models have thus found success in motion planning \citep{gutow2020koopman}, object tracking \citep{comas2021self}, robust control \citep{zhang2022robust}, and mobility applications \citep{manzoor2023vehicular}.

	Building upon the earlier work \citep{bueno2025koopman}, which first introduced the use of the Koopman operator for modeling and predicting unknown moving obstacles in single-agent UAV navigation, this paper advances the concept to a multi-agent cooperative setting with distributed and switched predictive control. The proposed framework unifies Koopman-based data-driven forecasting with distributed switching MPC (SMPC), enabling real-time, decentralized, and collision-free coordination among multiple UAVs in dynamic environments with moving or fixed obstacles.
	The contributions are: i) a distributed SMPC framework with flexible motion modes and clustering, scalable to large multi-vehicle networks and ii) a Koopman-based obstacle prediction model embedded in SMPC via linear constraints, enabling safety by avoiding dynamic agents and obstacles in real time.
	

	\noindent The remainder of the paper is structured as follows. Section~2 presents the problem formulation and preliminaries. Section~3 details the proposed method. Section~4 illustrates simulation results, and Section~5 concludes the paper with the proposal overview and suggests future directions.

	\section{Preliminaries and Problem Statement}
	    \label{Sec. 2}

	    This section details the quadrotor model, its reformulation into the switching framework with the corresponding discrete representation, and the problem statement.

	
	\subsection{Quadrotor UAV}
	This work considers a general UAV kinematic model suitable for various robotic platforms, assuming that embedded control loops handle the actuator dynamics. A local replanner is also assumed to ensure smooth velocity and acceleration references, including those from switched controllers (e.g., \citep{Replanning}).
	Let $O_{\text{W}} - X_{\text{W}} Y_{\text{W}} Z_{\text{W}}$ denote the fixed world frame and $O_{\text{B}}-X_{\text{B}} Y_{\text{B}} Z_{\text{B}}$ the body frame attached to each UAV. Using Euler’s equations, the absolute position and orientation vector $\bar{{p}} =[p \ \phi \ \theta \ \psi]^{\top}$, with $p = [p_x \ p_y \ p_z]^{\top}$, evolves based on the body velocity vector ${v}=[v_u \ v_v \ v_w \ w_p \ w_q \ w_r]^{\top}$ as
	\begin{flalign} \label{NONDYNINTRO}
	 \dot{\bar{{p}}} = R {v},
	\end{flalign}
	where $R=\text{diag}(R_v, R_w)$ is block diagonal with:
	
	\vspace{5pt}
	
	$R_v =
	\left[\begin{smallmatrix}
	c(\theta)c(\psi) & s(\phi)s(\theta)c(\psi)-c(\theta)s(\psi)& c(\phi)s(\theta)c(\psi)+s(\theta)s(\psi)\\
	c(\theta)s(\psi) & s(\phi)s(\theta)s(\psi)+c(\theta)c(\psi)& c(\phi)s(\theta)s(\psi)-s(\theta)c(\psi)\\
	-s(\theta) & s(\phi)c(\theta) & c(\phi)c(\theta)
	\end{smallmatrix}\right],$
	
	\vspace{8pt}
	
	$R_w = \left[\begin{smallmatrix}
	1 & s(\phi)\tan(\theta) & c(\phi)\tan(\theta)\\
	0 & c(\phi) & -s(\phi)\\
	0 & s(\phi)c(\theta)^{-1} & c(\phi)c(\theta)^{-1}
	\end{smallmatrix}\right]$.
	
	\vspace{5pt}
	
	\noindent
	Here, $c(\cdot)$ and $s(\cdot)$ denote $\cos(\cdot)$ and $\sin(\cdot)$. For clarity, we use the general continuous-time form of the kinematics in \eqref{NONDYNINTRO}, where the absolute UAV pose $\bar{{p}}$ is the state $x$ and the body velocity $v$ is the input $u$:
	\begin{flalign} \label{NONDYN000}
	\dot x = f(x,u),\quad x(0)=x_0.
	\end{flalign}

	
	\subsection{Switched UAV model}
	Switched systems offer a practical approach for large-scale distributed networks by reducing dynamics complexity through a finite set of motion modes \citep{dsmpc}. We recast \eqref{NONDYN000} into this framework.
	Consider $N_{\text{r}}$ identical quadrotors indexed by $\mathcal{N} \coloneqq \{i \mid i = 1, \dots, N_{\text{r}}\}$. Each UAV $^{[i]}$ operates under a mode $\sigma^{[i]} \in \mathcal{M} \coloneqq \{1, \dots, m\}$, governed by the control input $u^{[i]}(\sigma^{[i]}) \in \mathcal{U}_{\text{sw}}$, $\mathcal{U}_{\text{sw}}$ being a predefined set of velocity vectors:
	
	\begin{flalign} \label{SWITCHES}
	u^{[i]}(\sigma^{[i]})=\left\{
	\begin{array}{ll}
	0_{6\times1}, & \sigma^{[i]}=1\\
	\bar v e_{\sigma^{[i]}-1}, & \sigma^{[i]}=2,3,4\\
	\bar w e_{\sigma^{[i]}-1}, & \sigma^{[i]}=5,6,7\\
	-\bar v e_{\sigma^{[i]}-7}, & \sigma^{[i]}=8,9,10\\
	-\bar w e_{\sigma^{[i]}-7}, & \sigma^{[i]}=11,12,13\\
	\end{array}\right. .
	\end{flalign}
	
	Here, $\bar v$ and $\bar w$ are predefined velocities, and $e_{(\cdot)}$ denotes the canonical basis of $\mathbb{R}^6$. The continuous-time switched model for UAV $i$ becomes:
	\begin{flalign} \label{DYN000}
	\dot x^{[i]}= f_{\sigma^{[i]}}(x^{[i]},u^{[i]}(\sigma^{[i]})),\quad x^{[i]}(0) = x^{[i]}_{0}.
	\end{flalign}
	For implementation, we use the discrete version with sampling time $T > 0$ and the integer $k$ as sampling instant given by
	\begin{flalign} \label{DYN111}
	x^{[i]}_{k+1} &= x^{[i]}_{k}+T f^{[i]}_{\sigma_k}(x^{[i]}_k,u^{[i]}_{k}(\sigma^{[i]}_k)), \nonumber \\
	x^{[i]}_{k+1} &= {f_{\text{d}}}_{\sigma_k}(x^{[i]}_k,u^{[i]}_{k}(\sigma^{[i]}_k)).
	\end{flalign}
	\begin{figure}[t]
	\centering
	\Huge
	\begin{tikzpicture}[scale=0.47]
	\definecolor{combus}{rgb}{0.706,0.247,0.247}
	\definecolor{subsys}{rgb}{0.075,0.294,0.439}
	\filldraw[thick, rounded corners, color=subsys, fill=subsys!5]  
	                                (-0.5,3.75) rectangle(15.5,7.75)
	                                (-0.5,8.75) rectangle(15.5,12.75);
	
	\filldraw[thick, rounded corners, fill=white]  
	                                (2,5) rectangle(7,7)
	                                (2,10) rectangle(7,12);
	\filldraw[thick, rounded corners, fill=white]
	                                (10,4.5) rectangle(14,7.5)
	                                (10,9.5) rectangle(14,12.5);
	
	\draw[thick]    (4.5,8.45)node{\scriptsize $\vdots$}
	                (12,8.45)node{\scriptsize $\vdots$}
	                (4.5,6)node{\normalsize SMPC 1}
	                (4.5,11)node{\normalsize SMPC $N_{r}$};
	
	\node[inner sep=0pt] at (12,6) {
	\includegraphics[width=.08\textwidth]{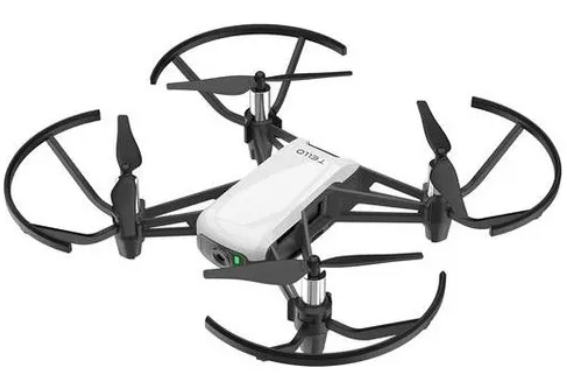}
	};
	\node[inner sep=0pt] at (12,11) {
	\includegraphics[width=.08\textwidth]{FIG-PDF/dron_1.PNG}
	};
	
	\draw[->,thick,dashed, color=combus] (0,4)--(0,13.5)node[right,yshift=0mm]{};
	\draw[->,thick,dashed, color=combus] (9,5.5)--(9,13.5)node[right,yshift=0mm]{};
	
	\draw[->,thick,dashed, color=combus] (0,6.5)--(1,6.5)node[above,yshift=-1.5mm,scale=0.8]{\scriptsize $\mathcal{X}^{[1]}_{\mathcal{C}_k}$}--(2,6.5);
	\draw[->,thick,dashed, color=combus] (0,11.5)--(1,11.5)node[above,yshift=-1.5mm,scale=0.8]{\scriptsize $\mathcal{X}^{[N_{r}]}_{\mathcal{C}_k}$}--(2,11.5);
	
	\draw[<->,thick,dashed, color=combus] (9,6.5)--(8,6.5)node[above,yshift=-1.5mm,scale=0.8]{\scriptsize $\mathcal{U}^{[1]}_{\mathcal{C}_k}$}--(7,6.5);
	\draw[<->,thick,dashed, color=combus] (9,11.5)--(8,11.5)node[above,yshift=-1.5mm,scale=0.8, color=combus]{\scriptsize $\mathcal{U}^{[N_r]}_{\mathcal{C}_k}$}--(7,11.5);
	
	\draw[->,thick] (7,5.5)--(8,5.5)node[above,yshift=-1.65mm,scale=0.95]{\scriptsize $u^{[1]}$}--(10,5.5);
	\draw[->,thick] (7,10.5)--(8,10.5)node[above,yshift=-1.65mm,scale=0.95]{\scriptsize $u^{[N]}$}--(10,10.5);
	
	\draw[->,thick] (14,6)--(15,6)node[above,yshift=-1.65mm,scale=0.95,xshift=-1.65mm]{\scriptsize $x^{[1]}$}--(16,6);
	\draw[->,thick] (14,11)--(15,11)node[above,yshift=-1.65mm,scale=0.95,xshift=-1.65mm]{\scriptsize $x^{[N_{r}]}$}--(16,11);
	
	\draw[->,thick] (15,6)--(15,4)--(1,4)--(1,5.5)--(2,5.5);
	\draw[->,thick] (15,11)--(15,9)--(1,9)--(1,10.5)--(2,10.5);
	
	\draw[thick] (1,4)--(0,4)
	            (1,9)--(0,9);
	\filldraw[thick, fill=combus, color=combus]    
	                    (0,4)circle[radius=3pt]
	                    (0,9)circle[radius=3pt]
	                    (9,5.5)circle[radius=3pt]
	                    (9,10.5)circle[radius=3pt];
	\end{tikzpicture}
	\caption{Sequential distributed SMPC scheme.}
	\label{fig:DMPC_arch}
	\end{figure}

	\subsection{Problem statement}
	Consider a closed 3D environment with $N_{\text{r}}$ identical quadrotor UAVs governed by \eqref{DYN111}, each equipped with localization sensors, and $N_{\text{obs}}$ uncontrolled moving obstacles. The objective is to design a distributed switched predictive control strategy that autonomously guides each UAV to its reference position while ensuring collision avoidance with both obstacles and fellow agents.

	\section{Main results}
	\label{Sec. 3}

	Managing multi-agent systems becomes computationally demanding as the number of subsystems increases, especially under centralized coordination. Therefore, to address this, we adopt a distributed communication scheme that distributes the computational load among agents.
	
	\noindent
	Following the sequential distributed strategy in \citep{dmpc_review}, each UAV locally solves an SMPC using a cluster-based topology. Spherical clusters with radius $r_{\text{cl}}$ are centered at each UAV’s reference position, and agents within the same cluster are considered neighbors. The cluster membership for UAV $i$ at time $k$ is defined by:
	\begin{flalign} \label{CLUSTER}
	\mathcal{C}^{[i]}_k &\coloneqq \{j \in \mathcal{N} \mid \| p^{[i]}_k - p^{[j]}_k \|_2 < r_{\text{cl}},\ j \ne i\}.
	\end{flalign}
	Letting $N$ be the horizon length, associated state and input information for the agents in $\mathcal{C}^{[i]}_k$ are collected in:
	\begin{flalign} \label{CLUSTER222}
	\mathcal{X}^{[i]}_{\mathcal{C}_k} &\coloneqq \{x^{[j]}_k \in \mathbb{R}^6 \mid j \in \mathcal{C}^{[i]}_k\}, \nonumber \\
	\mathcal{U}^{[i]}_{\mathcal{C}_k} &\coloneqq \{ \mathbf{u}^{[j]}_k = [\hat{u}^{[j]}_{k+1}, \dots, \hat{u}^{[j]}_{k+N-1}] \mid j \in \mathcal{C}^{[i]}_k\}.
	\end{flalign}
	To enhance robustness, we share only the optimal predicted input sequence $\mathcal{U}^{[i]}_{\mathcal{C}_k}$ instead of full predicted trajectories, avoiding noise propagation from state measurements.
	
	In this sequential setup, agents solve their SMPCs in a predefined order, each using updated input data from previously solved neighbors. At $k=0$, $\mathcal{U}^{[i]}_{\mathcal{C}_k}$ is empty; this is addressed either by initializing $\mathcal{C}^{[i]}_k = \emptyset$ or assuming neighbors remain static \citep{dsmpc}. This procedure is recursively applied to all agents yet to solve their optimal control problems. Fig.~\ref{fig:DMPC_arch} illustrates this distributed framework, showing subsystem clusters and inter-agent communication links.

	\begin{figure*}[!h]
	\centering
	\subfigure[Circular trajectory]{
	  \includegraphics[trim=0.0cm 0.0cm 0.0cm 0.0cm, clip, width=0.4\textwidth]{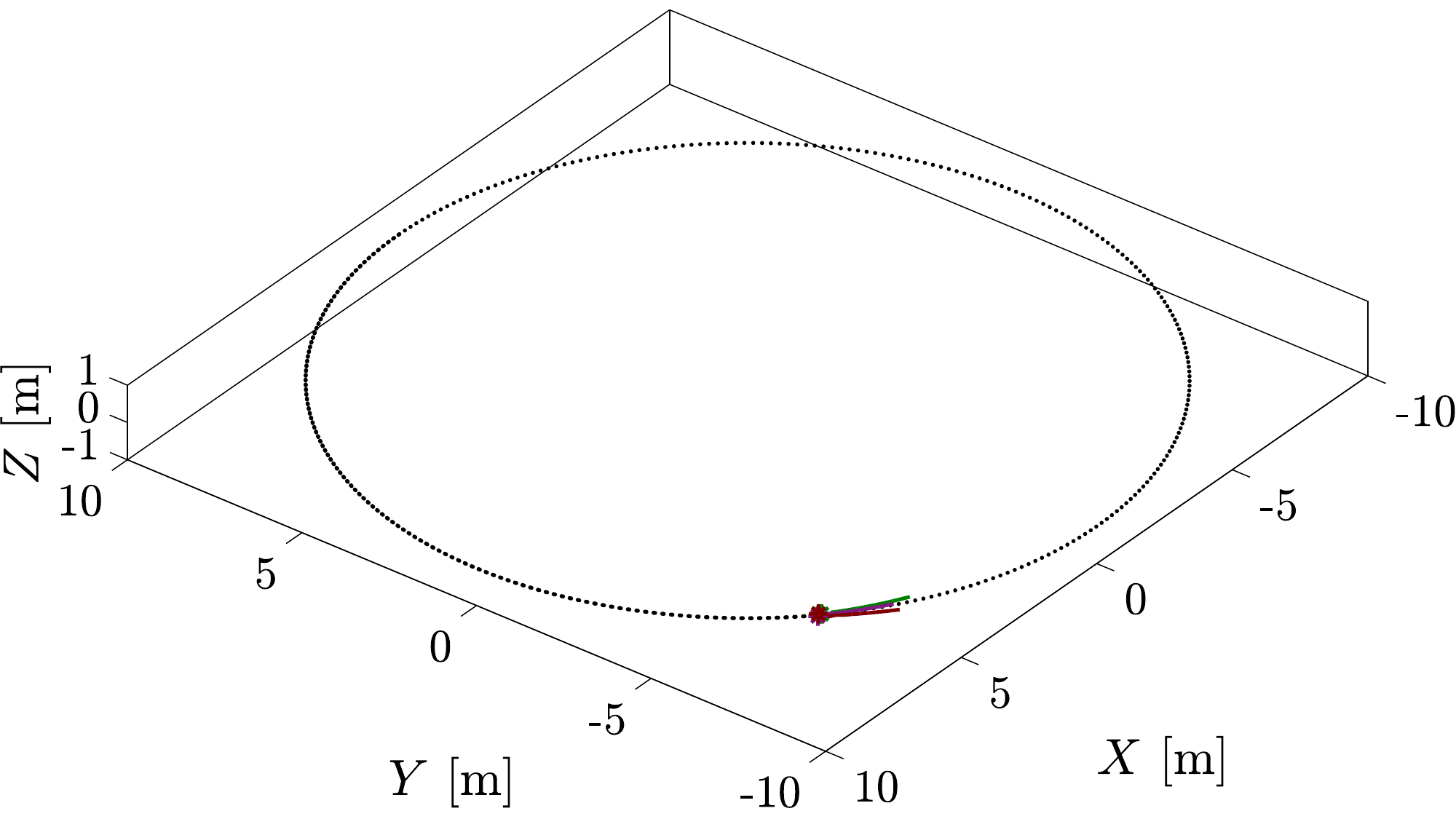}\label{fig:sim2xxxtrajjjj}
	}
	\subfigure[Horizontal figure-eight trajectory]{
	  \includegraphics[trim=0.0cm 0.0cm 0.0cm 0.0cm, clip, width=0.4\textwidth]{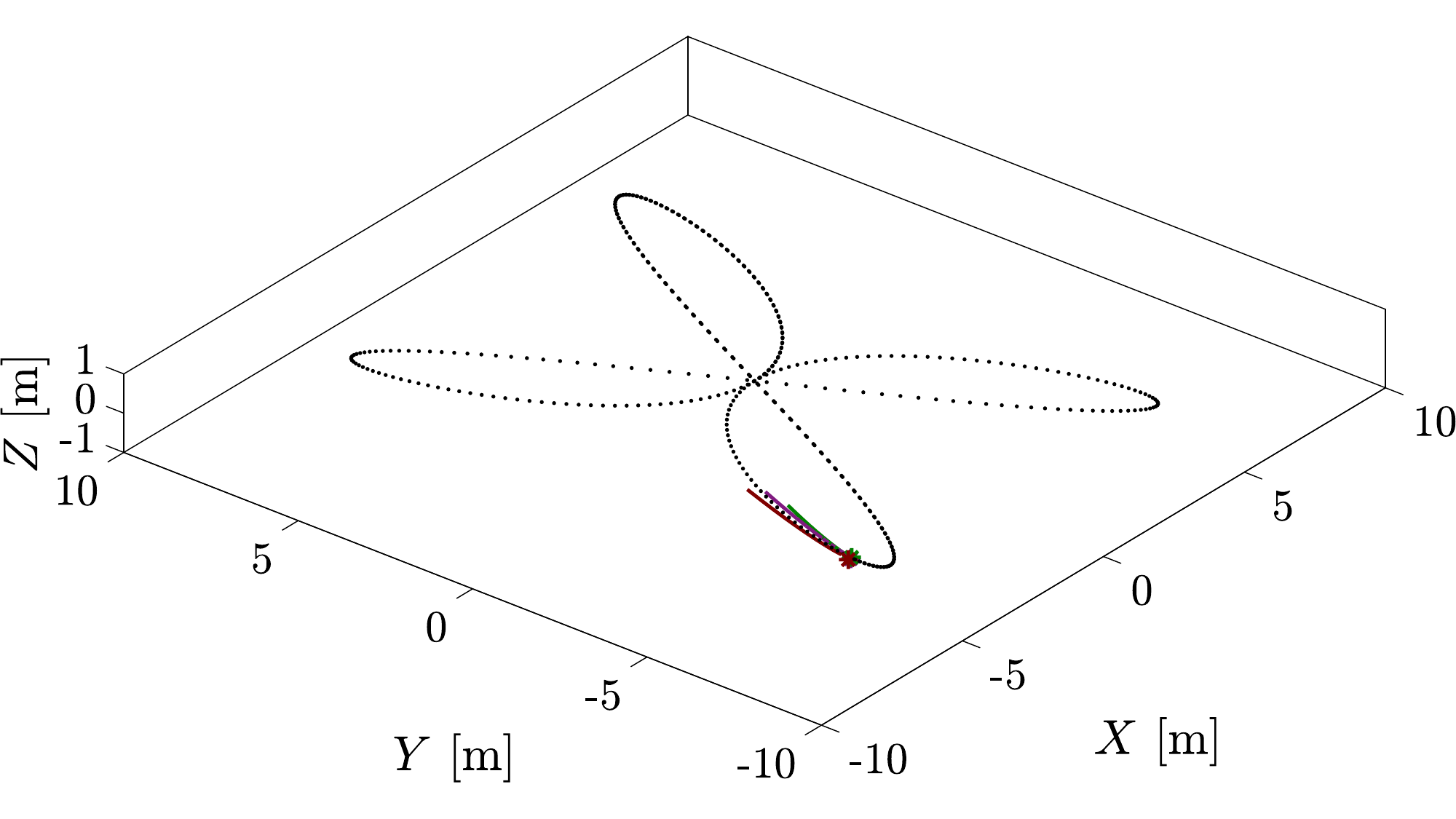}\label{fig:sim6xxxtrajjjj}
	}
	\captionsetup{font=footnotesize}
	\caption{Tracking and prediction of moving obstacles. 
	}
	\label{fig:snapsxxxKooppp}
	\end{figure*}

	\begin{figure}[b]
	\centering
	\includegraphics[trim=0.0cm 0.0cm 0.0cm 0.0cm, clip, width=0.4\textwidth]{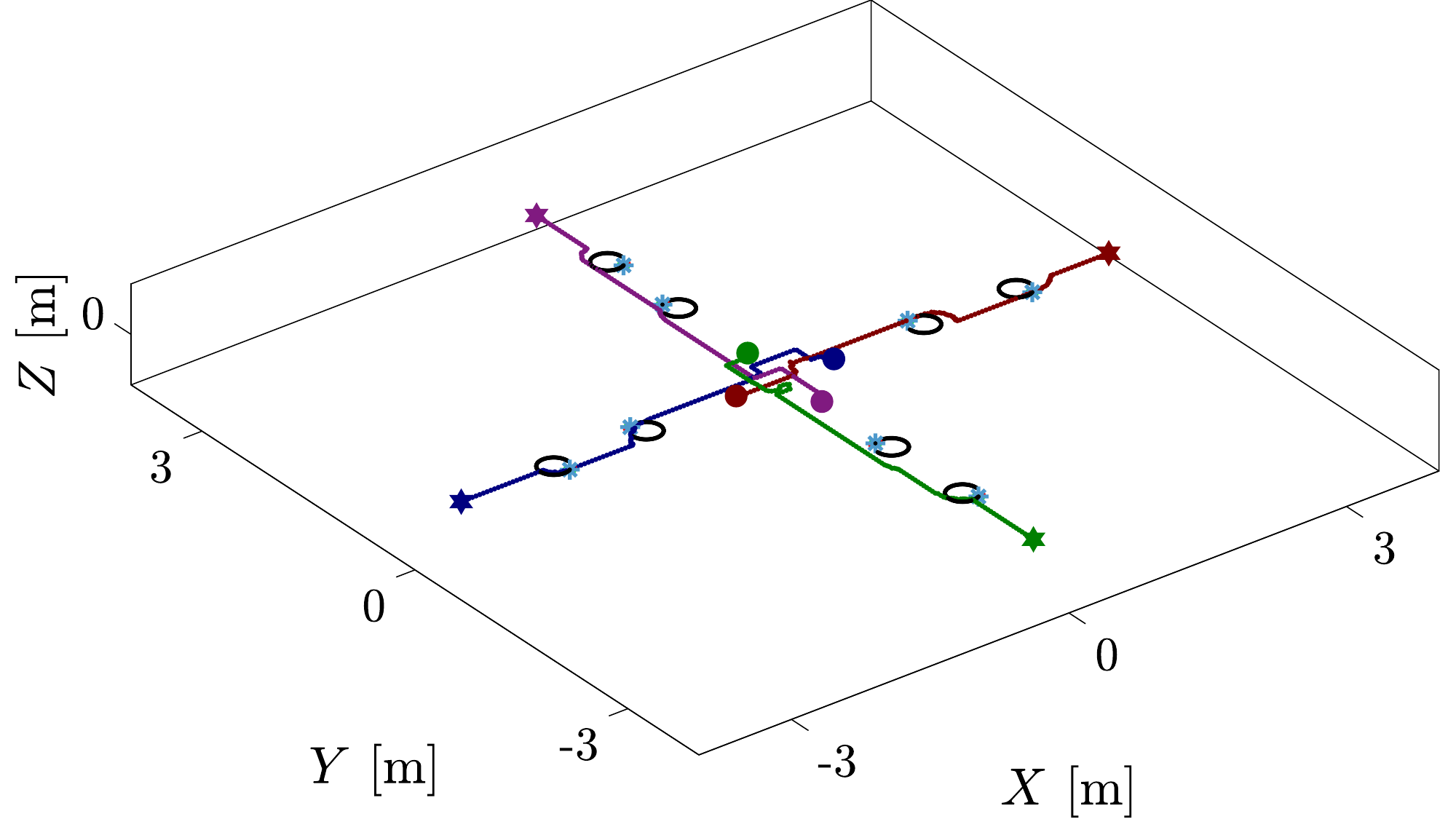}
	\captionsetup{font=footnotesize}
	\caption[]{Ultimate positions of obstacles and UAVs as well as their trajectories over time. 	\label{fig:sim}}
	\end{figure}

	\begin{table*}[!h]
	\centering
	\caption{Prediction error comparison.}
	\begin{tabular}{lcccccc}
	\toprule
	\multirow{2}{*}{Motion Type} & \multicolumn{3}{c}{Nonlinear GP \citep{olcay2024dynamic}} & \multicolumn{3}{c}{Koopman (Proposed)} \\
	\cmidrule(lr){2-4} \cmidrule(lr){5-7}
	& RMSE & MAE & MaxErr & RMSE & MAE & MaxErr \\
	\midrule
	Circular ($X-Y$ plane) & 64.2611 & 56.6285 & 110.9384 & 51.6039 & 48.4821 & 77.0244 \\
	Planar Figure-Eight Trajectory ($X-Y$ plane)  & 83.4153 & 72.8220 & 146.1566 & 69.2424 & 63.4870 & 108.2167 \\
	3-dimensional Butterfly Curve & 78.1790 & 68.8219 & 135.4376 & 39.5392 & 37.3492 & 53.0990 \\
	\bottomrule
	\end{tabular}
	\label{tab:error_comparison}
	\end{table*}

	\subsection{Koopman operator-based collision avoidance}
	Collision avoidance is essential for UAVs, particularly in environments with dynamic and unknown obstacles. This work addresses such challenges by predicting obstacle trajectories using the Koopman operator, which transforms nonlinear dynamics into a linear framework through lifted observables. 
	For a nonlinear discrete-time system $z(k+1) = f(z(k))$, with state $z(k) \in \R^n$, the Koopman operator $\mathcal{K}$ governs the evolution of observables $g(z)$ via
	\begin{flalign} \label{KOOPMAN}
	g(z(k+1)) = \mathcal{K} g(z(k)).
	\end{flalign}
    In the following, for the sake of simplicity, we adopt the notation $g_k(z_k)$ standing for $g(z(k))$.
	This enables linear prediction of complex dynamics using data-driven approximations \citep{williams2015data, zhang2019online}.
	Let $o^{[l_o]}_k \in \R^3$ be the estimated position of the $l_o$-th obstacle at time $k$. During each open-loop iteration, the lifted prediction evolves as \citep{bueno2025koopman}:
	\begin{flalign} \label{ObstacleDynamics}
	g_{t+1}(o_{t+1}^{[l_o]}) = \mathcal{K} g_{t}(o_t^{[l_o]}), ~~ \forall t\in\mathcal{T}_k,
	\end{flalign}
	with $\mathcal{T}_k = \{k, \dots, k+N-1\}$, with $N>0$ being the prediction horizon. The predicted position $o^{[l_o]}_t$ is extracted from $g_t(o_t^{[l_o]})$.
	These predicted positions are modeled as spherical obstacles. The collision constraint between UAV $i$ and obstacle $l_o$ is:
	\begin{flalign} \label{DistanceTo}
	    \| p^{[i]}_{t} - o^{[l_o]}_{t} \|_2 \geq R^{[i]} + R^{[l_o]} + \delta^{[i,l_o]},
	\end{flalign}
	where $R^{[i]}$ and $R^{[l_o]}$ denote safety radii, and $\delta^{[l_o]}$ is a safety margin. Due to its nonlinearity, we approximate the sphere by a polytope defined by $\gamma_t^{[i,l_o]}$ linear inequalities, i.e.,
	\begin{flalign} \label{polytope}
	    (\eta_{t,\mu}^{[i,l_o]})^{\top} (p^{[i]}_{t} - o^{[l_o]}_{t}) \leq d_{t}^{[i,l_o]}, \quad \mu = 1, \dots, \gamma_t^{[i,l_o]},
	\end{flalign}
	with $\eta_{t,\mu}^{[i,l_o]}$ as normal vectors and $d_{t}^{[i,l_o]}$ the corresponding plane distances.
	To enforce the safe distance, we compute the signed distances as
	\begin{flalign} \label{polytope2}
	    \rho_{t,\mu}^{[i,l_o]} = (\eta_{t,\mu}^{[i,l_o]})^{\top}(p^{[i]}_{t} - o^{[l_o]}_{t}) - d_{t}^{[i,l_o]},
	\end{flalign}
	and select the most restrictive plane with
	\begin{flalign} \label{polytopeMAX}
	    \mu_{t,\max}^{[i,l_o]} = \arg \max_{\mu} \rho_{t,\mu}^{[i,l_o]}.
	\end{flalign}
	The linear constraint ensuring avoidance becomes
	\begin{flalign} \label{LINEARCONSTRAINT}
	    (\eta_{t, \mu_{t,\max}^{[i,l_o]}}^{[i,l_o]})^{\top}(p^{[i]}_{t} - o^{[l_o]}_{t}) \geq d_{t}^{[i,l_o]}.
	\end{flalign}
	For inter-agent avoidance, a similar constraint applies, i.e.,
	\begin{flalign} \label{DistanceToAGENT}
	    \| p^{[i]}_{t} - p^{[j]}_{t} \|_2 \geq R^{[i]} + R^{[j]} + \delta^{[i,j]},
	\end{flalign}
	leading to the linearized version given by
	\begin{flalign} \label{LINEARCONSTRAINTINTERAGENT}
	    (\eta_{t, \mu_{t,\max}^{[i,j]}}^{[i,j]})^{\top}(p^{[i]}_{t} - p^{[j]}_{t}) \geq d_{t}^{[i,j]}.
	\end{flalign}
	These linear constraints ensure computational tractability and effective avoidance of both dynamic obstacles and other agents.

	\subsection{Finite horizon optimal control problem}
	Considering \eqref{SWITCHES}, \eqref{DYN111}, and linearized constraints \eqref{LINEARCONSTRAINT} and \eqref{LINEARCONSTRAINTINTERAGENT}, the optimal control problem for the $i$-th UAV is formulated as:
	\begin{subequations}
	\begin{flalign}
	\label{MPC_1}
	& \min_{\boldsymbol{\sigma}^{[i]}_k} \sum_{t\in\mathcal{T}_k} \Vert p^{[i]}_t - p_{\text{ref}}^{[i]} \Vert_2, \\
	& \text{s.t.} \ \forall t\in\mathcal{T}_k,~ j\in\mathcal{C}^{[i]}_k,~ j\ne i,~ l = 1, \dots, N_{\text{obs}}, \nonumber \\
	\label{MPC_2}
	& x^{[i]}_{t+1} = f^{\text{d}}_{\sigma_t^{[i]}}(x^{[i]}_t, u^{[i]}_t(\sigma^{[i]}_t)), \quad x^{[i]}_k = \tilde{x}^{[i]}_k, \\
	\label{MPC_3}
	& u^{[i]}_t \in \mathcal{U}_{\text{sw}}, \\
	\label{MPC_5}
	& (\eta_{t, \mu_{t,\max}^{[i,l_o]}}^{[i,l_o]})^{\top}(p^{[i]}_t - o^{[l_o]}_t) \geq d_t^{[i,l_o]}, \\
	\label{MPC_4}
	& (\eta_{t, \mu_{t,\max}^{[i,j]}}^{[i,j]})^{\top}(p^{[i]}_t - p^{[j]}_t) \geq d_t^{[i,j]}.
	\end{flalign}
	\end{subequations}
	Here, $\tilde{x}^{[i]}_k$ denotes the current state measurement. The cost \eqref{MPC_1} penalizes deviation from the reference position $p_{\text{ref}}^{[i]}$ across the prediction horizon $N$. The Koopman-based obstacle prediction informs constraint \eqref{MPC_5}, while inter-agent safety is enforced through \eqref{MPC_4}.
	\begin{figure*}
	\centering
	\subfigure[$t = 7$ sec.]{
	  \includegraphics[trim=6.0cm 0.0cm 7.0cm 0.0cm, clip, width=0.23\textwidth]{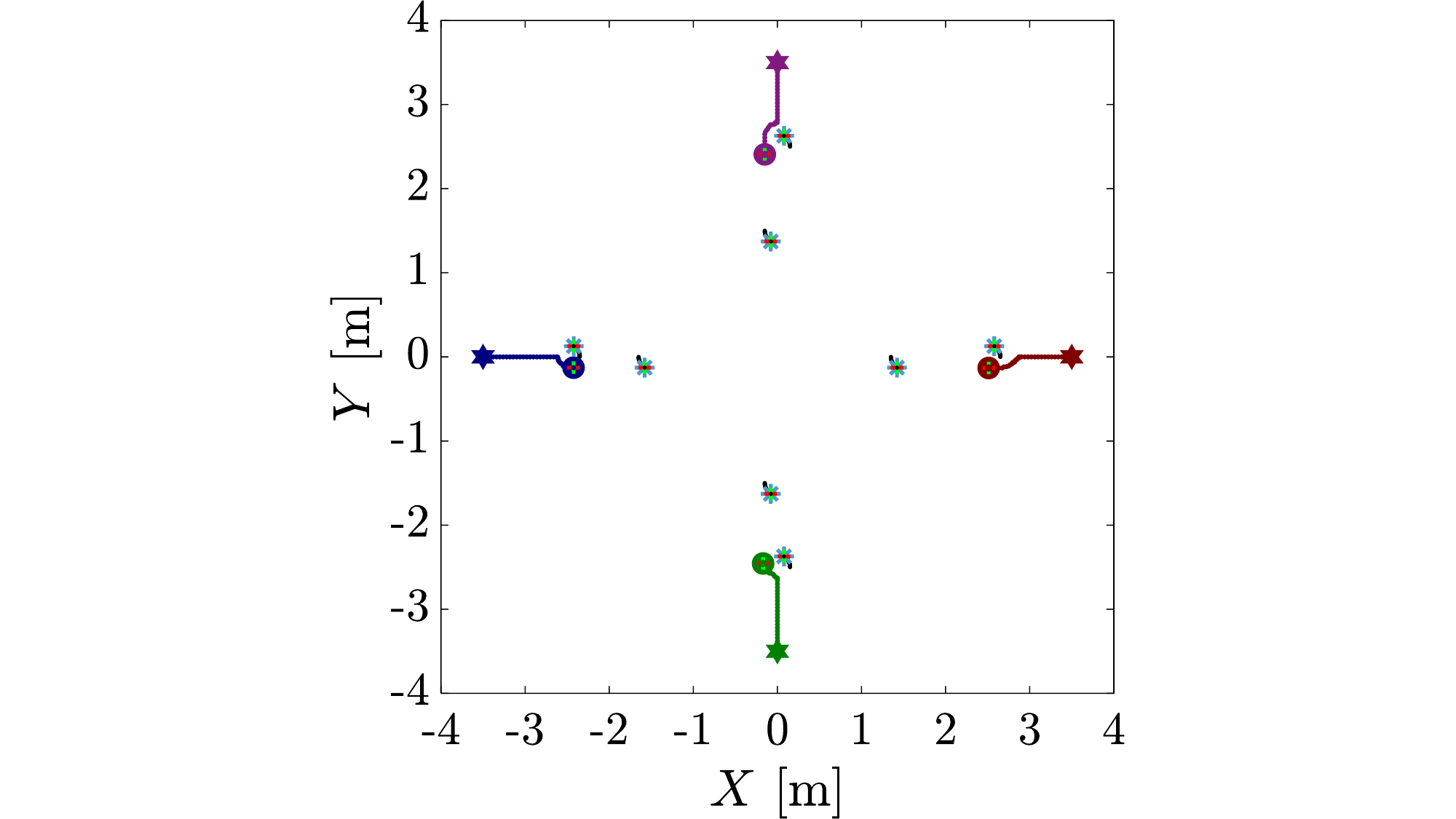}\label{fig:sim22xxx}
	}
	\subfigure[$t = 9$ sec.]{
	  \includegraphics[trim=6.0cm 0.0cm 7.0cm 0.0cm, clip, width=0.23\textwidth]{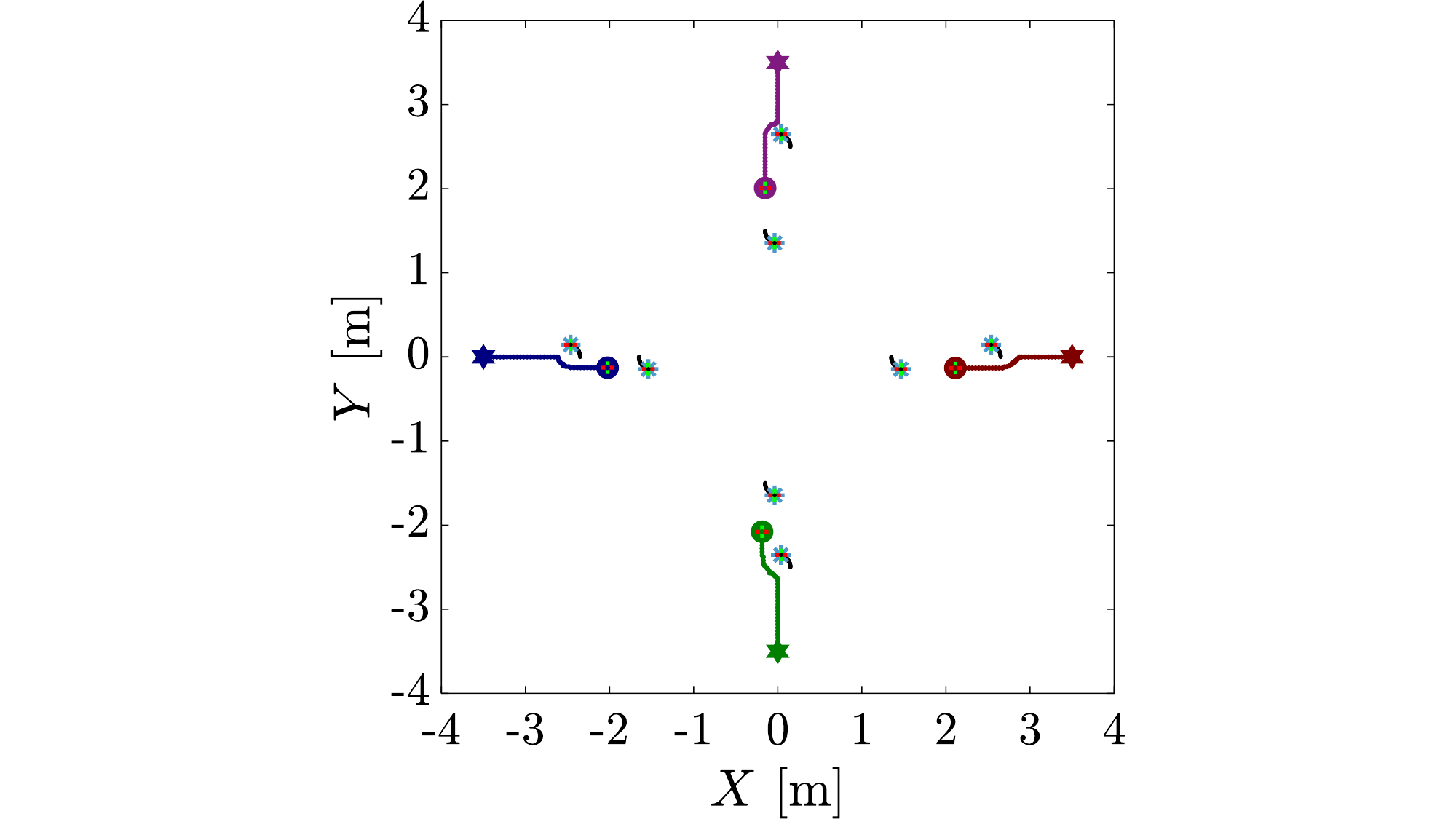}\label{fig:sim33xxx}
	}
	\subfigure[$t = 11$ sec.]{
	  \includegraphics[trim=6.0cm 0.0cm 7.0cm 0.0cm, clip, width=0.23\textwidth]{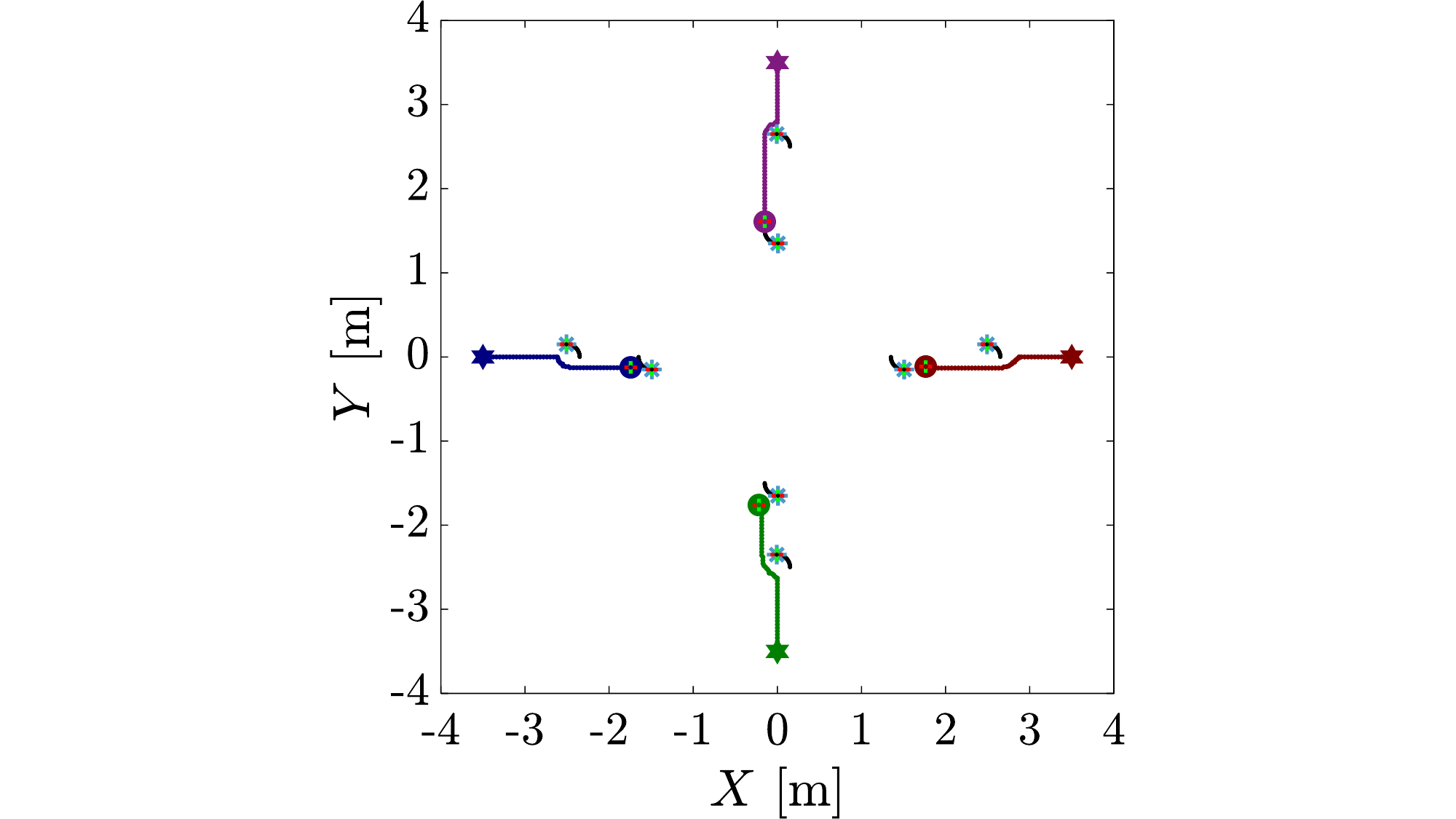}\label{fig:sim44xxx}
	}
	\subfigure[$t = 13$ sec.]{
	  \includegraphics[trim=6.0cm 0.0cm 7.0cm 0.0cm, clip, width=0.23\textwidth]{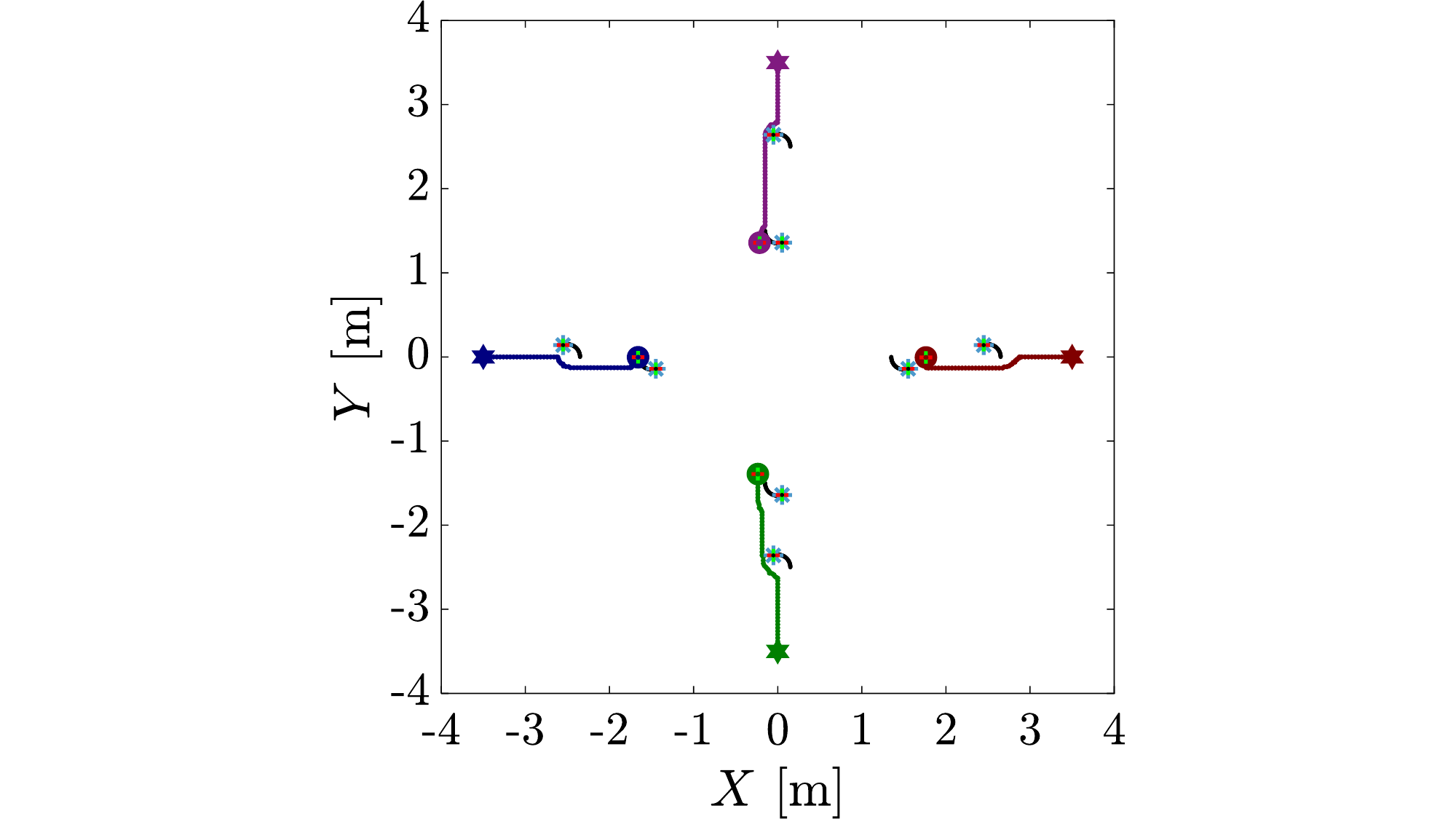}\label{fig:sim55xxx}
	}
	\captionsetup{font=footnotesize}
	\caption{Snapshots of UAVs and obstacles positions at different time instances.}
	\label{fig:snapsxxx}
	\end{figure*}
	\begin{figure*}[t]
	\centering
	\subfigure[UAV 1 and obstacles]{
	\includegraphics[trim=0.0cm 1.5cm 0.0cm 1.5cm, clip, width=0.4\textwidth]{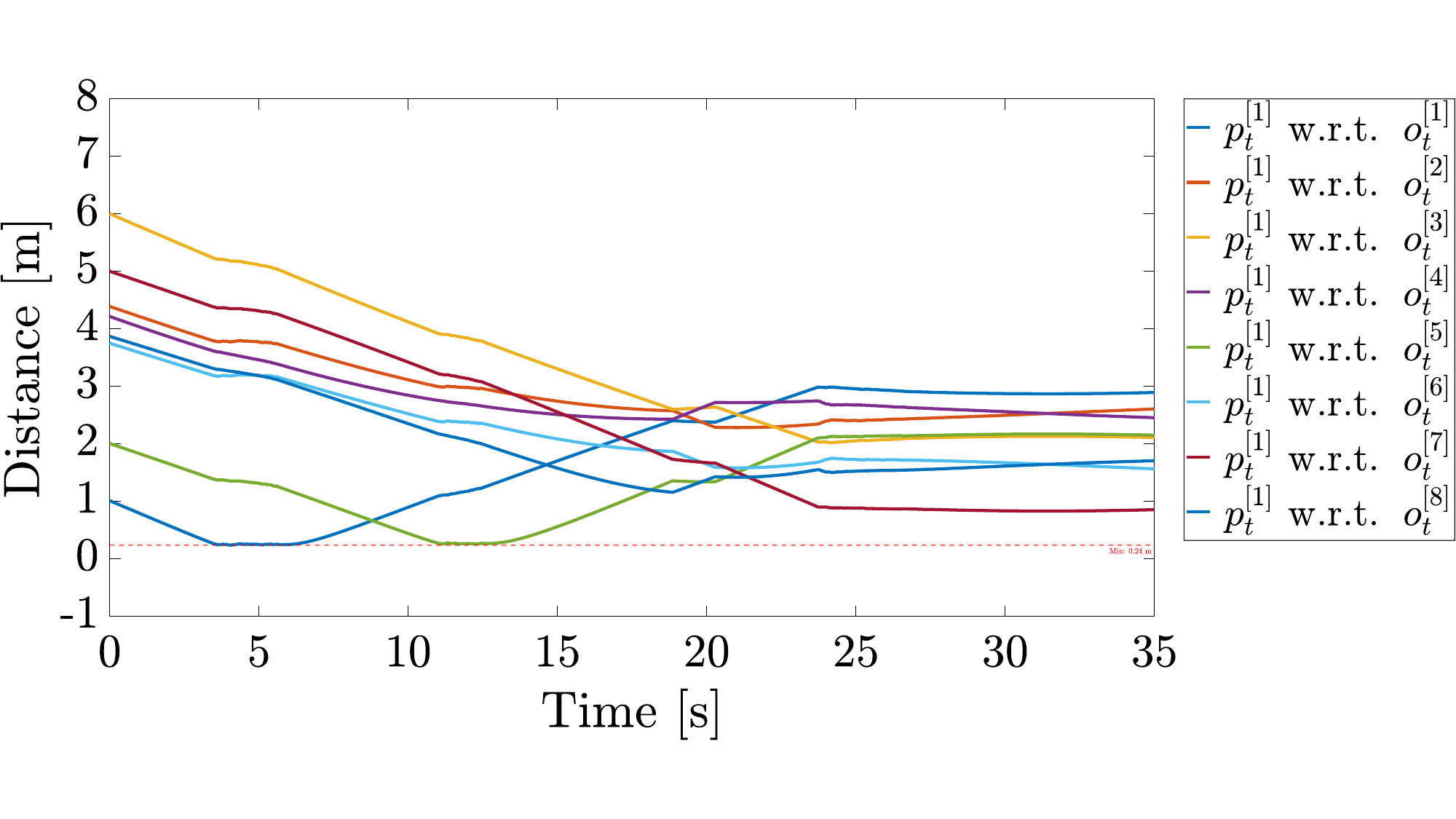}\label{fig:sim11xxxdist}
	}
	\subfigure[UAV 2 and obstacles]{
	\includegraphics[trim=0.0cm 1.5cm 0.0cm 1.5cm, clip, width=0.4\textwidth]{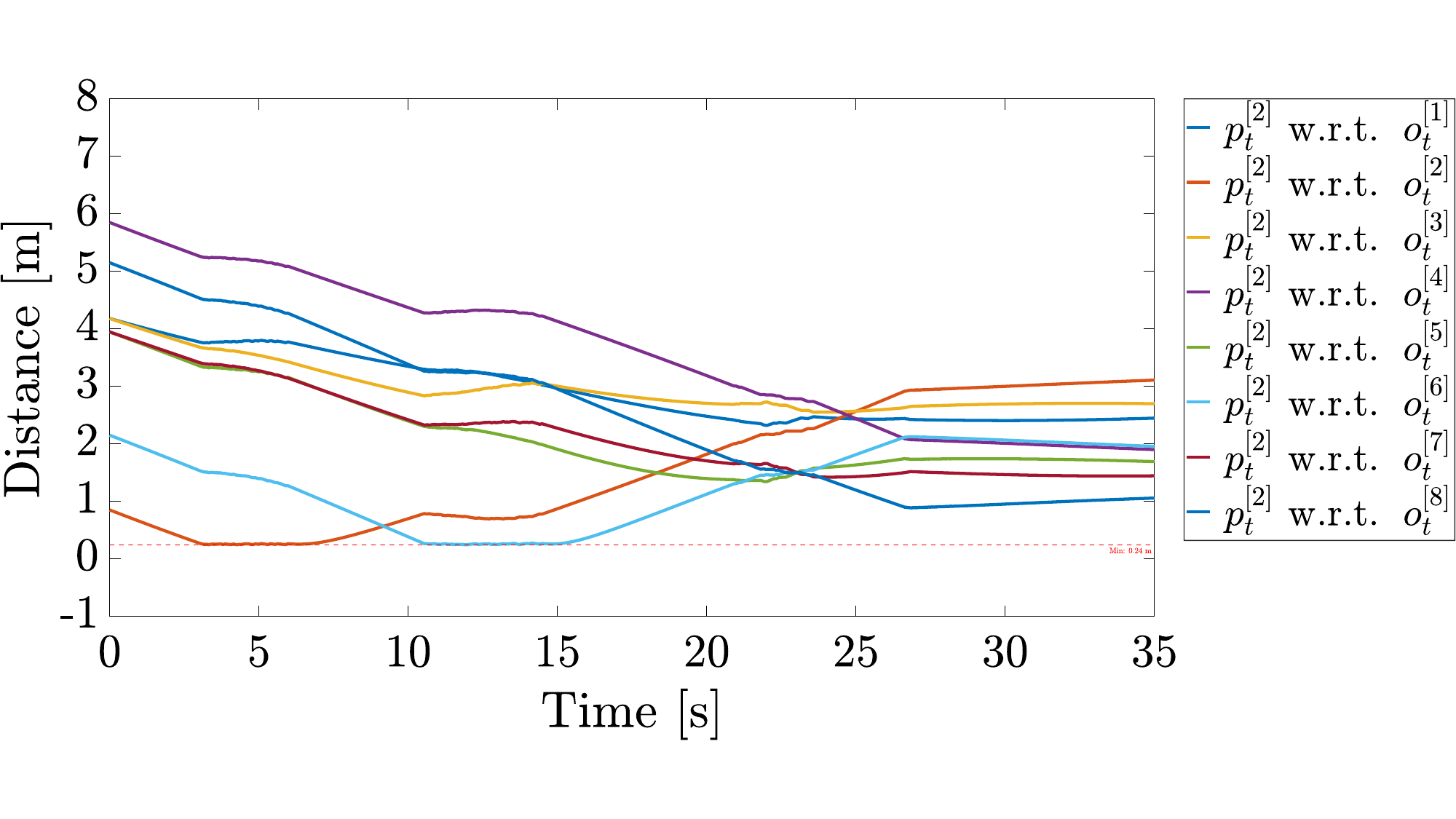}\label{fig:sim22xxxdist}
	}
	\subfigure[UAV 3 and obstacles]{
	\includegraphics[trim=0.0cm 1.5cm 0.0cm 1.5cm, clip, width=0.4\textwidth]{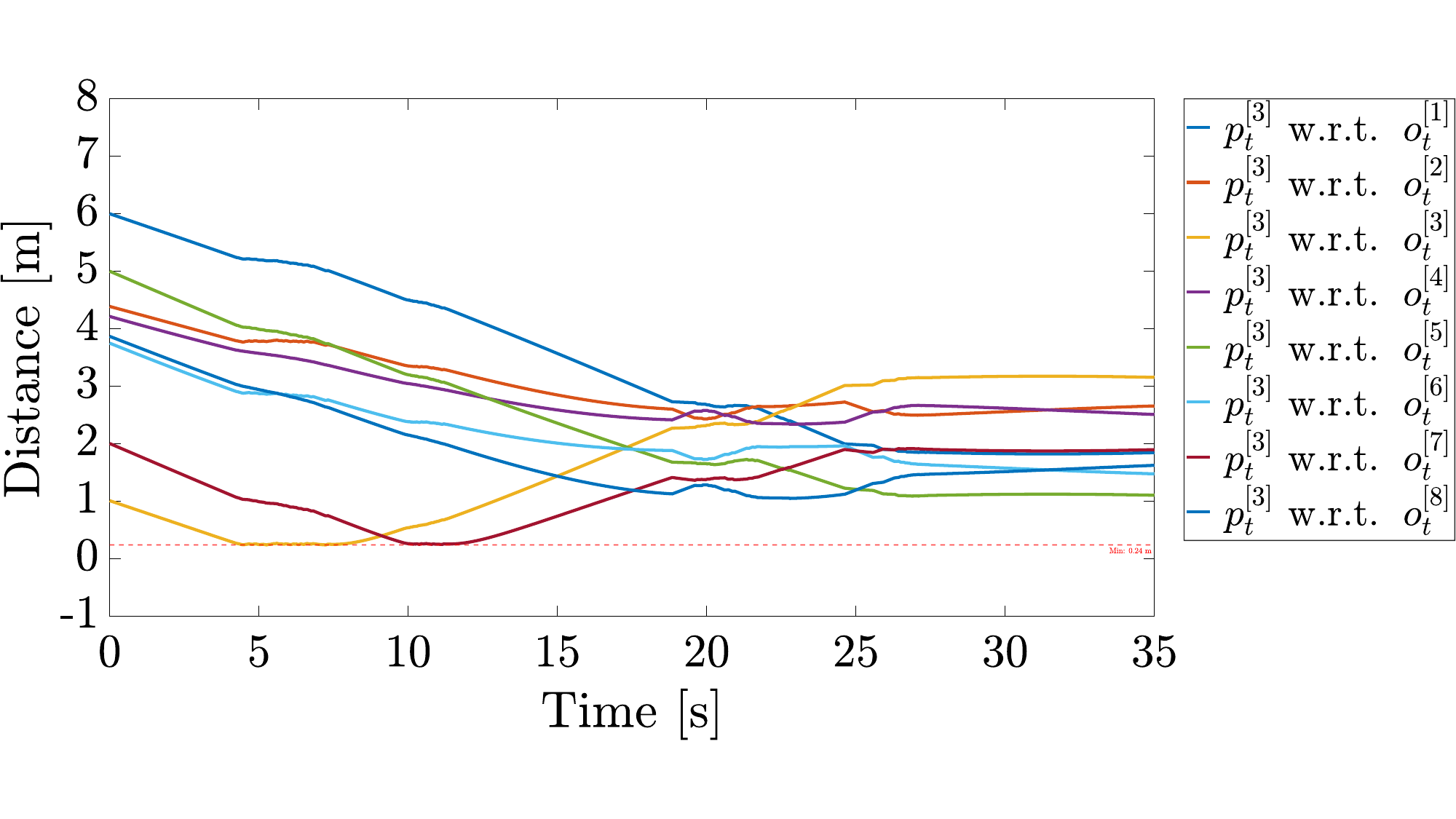}\label{fig:sim33xxxdist}
	}
	\subfigure[UAV 4 and obstacles]{
	\includegraphics[trim=0.0cm 1.5cm 0.0cm 1.5cm, clip, width=0.4\textwidth]{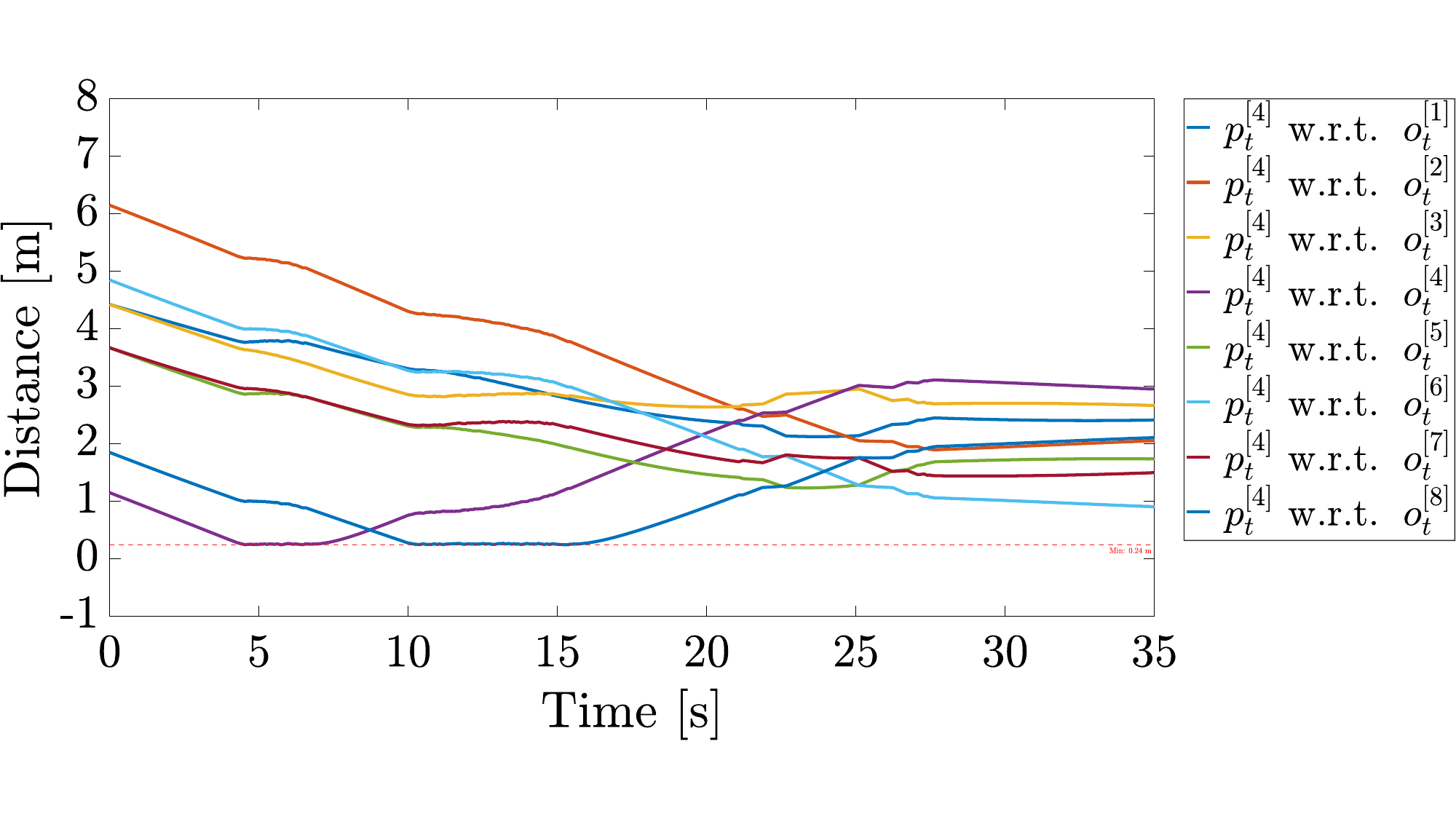}\label{fig:sim44xxxdist}
	}
	\captionsetup{font=footnotesize}
	\caption{Evolution of Euclidean UAV-to-obstacle distances. Highlighted minima for safe separation is $\mathrm{0.24~m}$.}
	\label{fig:Distance}
	\end{figure*}
	
	\section{Simulation results}
	\label{Sec. 7}
	 The simulation results are presented in two parts to highlight the effectiveness of the proposed Koopman-based modeling and prediction, as well as its integration with the SMPC framework for dynamic obstacle avoidance.

	\subsection{Koopman-based prediction of moving objects}
	In the first scenario, $N_{\text{r}}=3$ UAVs monitor a ground-moving obstacle without executing the SMPC task, emphasizing the Koopman-based approach’s ability to model and predict unknown dynamics over time. Sensor noise is applied based on the UAV-obstacle distance, realistically simulating measurement accuracy degradation with range. Once an obstacle enters the sensor range, it is continuously tracked, and position measurements are stored in a finite buffer. Upon collecting enough data, the Koopman operator $\mathcal{K}$ is computed \citep{bueno2025koopman}, yielding a linear approximation of the obstacle's behavior.
	To represent the motion, the lifting function is defined as:
	\begin{align*}
	g(o^{[l_o]}) &= [g(o_x^{[l_o]})^{\top} , g(o_y^{[l_o]})^{\top}]^{\top}, \\
	    g(o_\zeta^{[l_o]}) &= [ \zeta, \zeta^2, \sin(\zeta), \cos(\zeta), \zeta^2\sin(\zeta), \zeta^2\cos(\zeta)]^{\top},
	\end{align*}
	where $\zeta\in\{x^{[l_o]},y^{[l_o]}\}$ are the measured positions of the $l_o$-th obstacle. This formulation captures diverse dynamics by incorporating polynomial and trigonometric interactions, enabling the modeling of both smooth and complex nonlinear behaviors.
	Two distinct obstacle motions are considered: a circular trajectory in the $X-Y$ plane and a horizontal figure-eight path. The former reflects structured periodic motion typical in surveillance or patrol tasks, while the latter introduces variable curvature and speed, representing more complex adaptive behaviors. These trajectories offer a realistic benchmark for evaluating prediction performance in dynamic environments.
	Using the most recent observation, the Koopman model iteratively forecasts future positions over a finite horizon. As shown in Fig.~\ref{fig:snapsxxxKooppp} for a specific sample time, predicted trajectories (solid lines) closely follow the actual path (dotted line), even with distance-dependent measurement noise. This highlights the Koopman approach’s robustness and its utility in enabling proactive, collision-free path planning in the full control scenario.

	\subsection{Comparison and discussion}
	To validate the proposed obstacle prediction approach, we evaluate three standard metrics: root mean square error (RMSE), mean absolute error (MAE), and maximum absolute error (MaxErr). The Koopman-based method is compared against the nonlinear Gaussian process (GP) regression approach from~\citep{olcay2024dynamic}, which employs a custom basis function and a squared exponential kernel.
	Table~\ref{tab:error_comparison} summarizes the accumulated prediction errors computed over time by the three UAVs predicting obstacle motion online across three motion patterns. In all cases, the Koopman model outperforms the GP-based approach. For circular motion, RMSE drops from 64.26 to 51.60 and MaxErr from 110.94 to 77.02. In the planar figure-eight case, RMSE decreases from 83.41 to 69.24, with notable MAE and MaxErr improvements. The most significant gains appear in the 3-dimensional butterfly trajectory, where RMSE and MaxErr reduce from 78.18 and 135.44 to 39.54 and 53.10, respectively. These results demonstrate the Koopman model’s superior generalization and robustness in nonlinear motion prediction.
	\begin{figure}[b]
	\centering
	\includegraphics[trim=0.0cm 1.5cm 0.0cm 1.5cm, clip, width=0.4\textwidth]{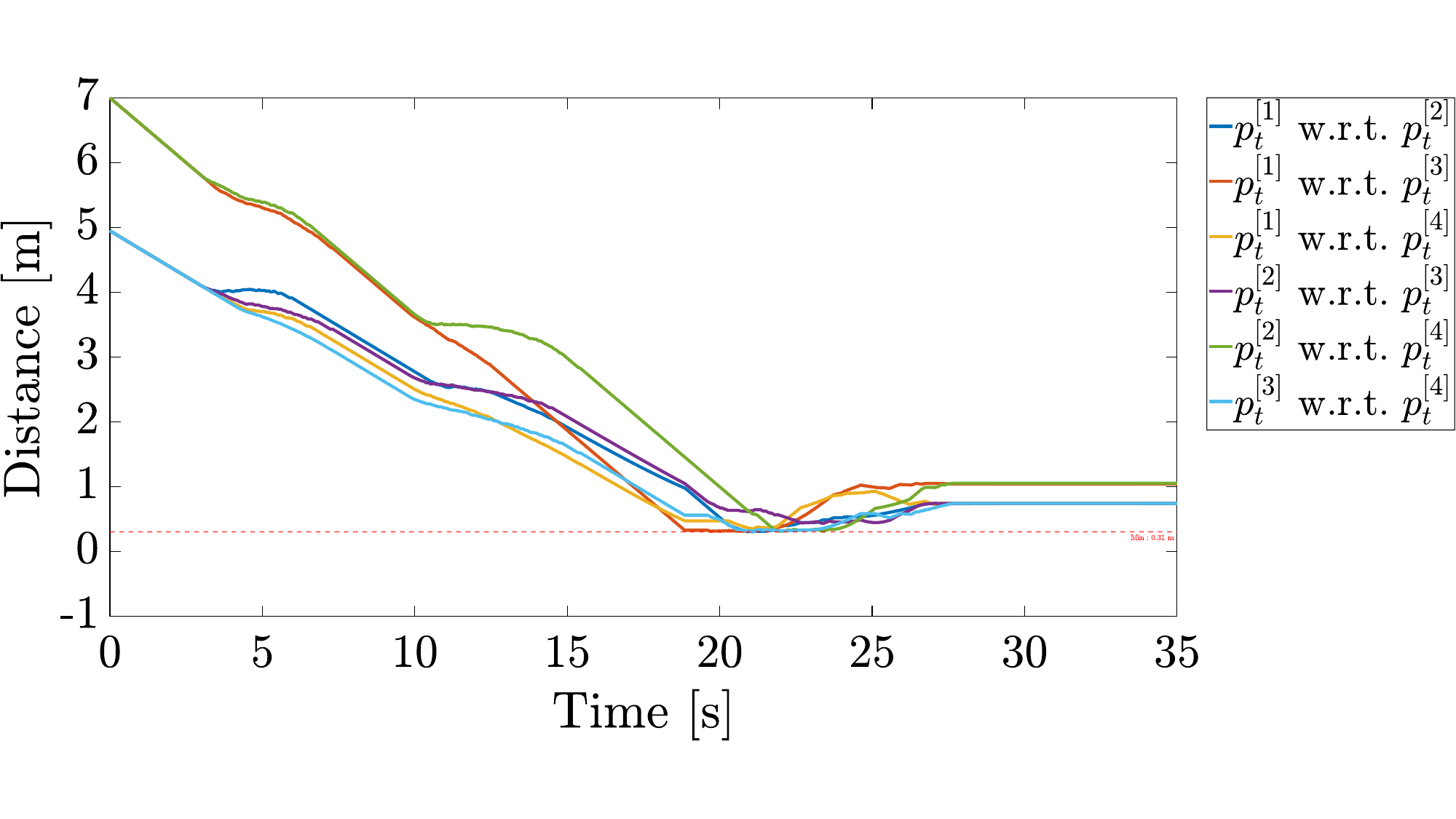}
	\captionsetup{font=footnotesize}
	\caption[]{Evolution of Euclidean inter-agent distances. Highlighted minima is $\mathrm{ 0.31~[m]}$.	\label{fig:sim55xxxdist}}
	\end{figure}

	\subsection{SMPC with dynamic obstacle avoidance}
	In the second scenario, UAVs perform obstacle avoidance using the SMPC framework while leveraging the Koopman-based predictions. The simulation uses a sampling time $T=0.01\ \text{s}$ and horizon $N=4$, over a 35-second window. UAVs and obstacles share the same physical parameters, including a frame radius $r_{\text{rob}}=r_{\text{obs}}=0.1125\ \text{m}$, used to define encapsulating spheres and tuning parameters.
	A team of $N_{\text{r}}=4$ controllable UAVs with dynamics \eqref{DYN111} and predefined velocities $\bar v=0.2\ \text{m}.\text{s}^{-1}$, $\bar w=0.6\ \text{rad}.\text{s}^{-1}$ begins at
	\begin{equation}
	\begin{aligned}
	x^{[i]}(0)=r_{\text{init}}[\cos{\dfrac{2 \pi (i-1)}{N_{\text{r}}}}, \sin{\dfrac{2 \pi (i-1)}{N_{\text{r}}}}, \bm{0}_4^{\top}]^{\top},
	\end{aligned}
	\label{eq:dispertion_metric}
	\end{equation}
	with $r_{\text{init}}=3.5\ \text{m}$ and $i=1,\dots,N_{\text{r}}$. Communication is allowed within clusters of radius $r_{\text{cl}}=8r_{\text{rob}}$.
	We deploy $N_{\text{obs}}=8$ uncontrolled obstacles arranged in two groups. Each obstacle's position is given by
	\begin{equation}
	\begin{aligned}
	&x^{[l,q]}_{\text{ring}}=r^{[q]}_{\text{ring}}[\cos{\dfrac{4 \pi (l-1)}{N_{\text{obs}}}},\ \sin{\dfrac{4 \pi (l-1)}{N_{\text{obs}}}},\ 0]^{\top},\\
	&x^{[l,q]}_{\text{cov}}=r^{[q]}_{\text{cov}}[\cos{\alpha^{[l,q]}_t},\ \sin{\alpha^{[l,q]}_t},\ 0]^{\top},\\
	&x^{[l,q]}_{\text{obs}}=x^{[l,q]}_{\text{ring}}+x^{[l,q]}_{\text{cov}},
	\end{aligned}
	\label{eq:dispertion_metric}
	\end{equation}
	where $l=1,\dots,4$ is the index of obstacles inside a $q$-th  ring, $q=1,2$, with ring radii equal to $r^{[1]}_{\text{ring}}=2.5\ \text{m}$, $r^{[2]}_{\text{ring}}=1.5\ \text{m}$, $r_{\text{cov}}=0.3\ \text{m}$ is the radius of the covered path, and $\alpha^{[l,q]}_t$ is the angular position of the obstacle along the covered path, with angular velocity chosen as 0.15 rad/s.
	The reference positions lie at the vertices of a square centered at the origin, i.e.,
	\begin{equation}
	\begin{aligned}
	x^{[i]}_{\text{ref}}=r_{\text{ref}}[-\cos{\dfrac{2 \pi (i-1)}{N_{\text{r}}}}, -\sin{\dfrac{2 \pi (i-1)}{N_{\text{r}}}}, \bm{0}_4^{\top}]^{\top},
	\end{aligned}
	\label{eq:dispertion_metric}
	\end{equation}
	with $r_{\text{ref}}=0.5\ \text{m}$ and $i=1,\dots,N_{\text{r}}$.
	
	Since UAVs are actively navigating, obstacle center of mass positions are measured by fixed ceiling cameras and relayed in real time. 
	In the simulation all motion occurs in the $X-Y$ plane, avoiding $Z$-direction maneuvers, thus enforcing challenging coordinated responses to dynamic obstacles which are constrained to the 2-dimensional domain. This setup highlights the true capability of the Koopman-based prediction and switching control strategy.
	
	Fig.~\ref{fig:sim} shows UAV trajectories navigating a dynamic environment with moving obstacles, culminating in successful formation while avoiding collisions. The Koopman-based prediction integrated with SMPC enables real-time, coordinated, and safe navigation to the target configuration.
	To further illustrate system behavior, Fig.~\ref{fig:snapsxxx} provides snapshots at selected time steps, showing UAV positions relative to obstacles. These frames reveal how agents dynamically adjust paths to avoid collisions while maintaining formation.
	Safety is assessed in Figs.~\ref{fig:sim11xxxdist}-\ref{fig:sim44xxxdist} and Fig.~\ref{fig:sim55xxxdist}, which display UAV-obstacle and inter-agent distances over time, with minimum distances highlighted. All separations remain above twice the UAV radius, confirming collision-free operation. This validates the effectiveness of our Koopman-SMPC framework in maintaining safety margins during navigation.

	\section{Conclusions}
	\label{Sec. 5}
	
	This paper introduced a Koopman-integrated distributed SMPC framework for cooperative UAV navigation in dynamic environments. The switching-based control reduces computation while retaining responsive planning. Leveraging the Koopman operator enables efficient  prediction of obstacles prediction. The distributed design enhances scalability and autonomy, supporting UAV traffic management. Simulations validate safe, collision-free navigation with preserved formation, showing promising results for aerial and surface mobility. Future works include extending to heterogeneous agents, aerial-ground coordination, and learning-based switching for dynamic environments.

	\bibliography{biblio}
	\end{document}